\documentclass[aps,nofootinbib,nobibnotes,noeprint,preprintnumbers,floatfix,twocolumn]{revtex4-2}
\usepackage{graphicx} % Required for inserting images
\usepackage{hyperref}
\usepackage[utf8]{inputenc}
\usepackage{graphicx}
\usepackage{amsmath}
\usepackage{xcolor}
\usepackage[caption = false]{subfig}
\usepackage{graphicx}% Include figure files
\usepackage{dcolumn}% Align table columns on the decimal point
\usepackage{bm}% bold math
\usepackage{hyperref}
\usepackage{xcolor}
\usepackage{amsmath,amsfonts,amssymb}
\usepackage{commath}
\usepackage{mathtools}
\usepackage{tikz}
\usepackage[utf8]{inputenc} 
\usepackage[T1]{fontenc}

\usepackage{graphicx}	% Including figure files
\usepackage{amsmath}	% Advanced maths commands
\newcommand{\be}{\begin{equation}} 
\newcommand{\ee}{\end{equation}}
\newcommand{\bea}{\begin{equation}\begin{aligned}} 
\newcommand{\eea}{\end{aligned}\end{equation}}

\newcommand{\msun}{{\,\rm M_\odot}}

\newcommand{\kms}{\,{\rm km}\,{\rm s}^{-1}}

\newcommand{\Gyr}{\,{\rm Gyr}}

\newcommand{\Myr}{\,{\rm Myr}}

\newcommand{\cpm}{\,{\rm cm}^2\,{\rm g}^{-1}}

\begin{document}
\preprint{FERMILAB-PUB-25-0113-T}
\title{Massive Black Holes Seeded by Dark Matter -- Implications for Little Red Dots and Gravitational Wave Signatures}
\author{Tingwei Shen$^{1}$}
\email{shencharline8@gmail.com}
\author{Xuejian Shen$^{2}$}
\email{xuejian@mit.edu}
\author{Huangyu Xiao$^{3,4}$}
\email{huangyu@fnal.gov}
\author{Mark Vogelsberger$^{2}$}
\author{Fangzhou Jiang$^5$}
%\author{Others}
\vspace{0.4cm}
\affiliation{$^1$ John A. Paulson School of Engineering and Applied Sciences, Harvard University, Boston, MA 02134, USA}
\affiliation{$^2$ Kavli Institute for Astrophysics and Space Research, Massachusetts Institute of Technology, Cambridge, MA 02139, USA}
\affiliation{$^3$ Theory Division, Fermilab, Batavia, IL 60510, USA}
\affiliation{$^4$ Kavli Institute for Cosmological Physics, University of Chicago, Chicago, IL 60637, USA}
\affiliation{$^5$ Kavli Institute for Astronomy and Astrophysics, Peking University, Beijing 100871, China}

\begin{abstract}
Observations of supermassive black holes (SMBHs) at high redshifts challenge standard seeding scenarios. We examine a dissipative self-interacting dark matter (dSIDM) model in which gravothermal collapse leads to the formation of massive BH seeds ab initio. We utilize a semi-analytical framework to predict properties of the dSIDM-seeded SMBH population. Billion solar mass quasars are reproduced along with low-mass faint active galactic nuclei (known as little red dots) with SMBH-to-galaxy stellar mass ratios consistent with recent James Webb Space Telescope observations. To match the abundance of the observed bright quasars, a sub-percent-level duty-cycle is suggested, implying a large population of dormant SMBHs. The gravitational wave (GW) signals from mergers of these massive SMBHs can be detected by LISA while remaining within the NANOGrav constraints on the GW background. These results provide testable signatures of DM-driven SMBH formation, offering a pathway to probe hidden-sector physics through SMBH and GW observables.
\end{abstract}
\maketitle

\section{Introduction}

The seeding and growth mechanisms of supermassive black holes (SMBHs) at high redshifts remain an open question (see e.g. the review \cite{Inayoshi2020}). One puzzle comes from observations of bright quasars at $z\gtrsim 6$, indicating that SMBHs with masses greater than a billion solar mass had already formed when the Universe was only $\sim 1\Gyr$ old~\citep[e.g.][]{Mortlock2011,Venemans2013,Wu2015,Mazzucchelli2017,Banados2018,Matsuoka2019,Yang2020,Wang2021}. This rapid growth requires either initially heavy BH seeds or sustained super-Eddington accretion \citep[e.g.][]{Madau2001,Bromm:2002hb,Schneider2002,Koushiappas:2003zn,Begelman:2006db,Lodato:2006hw,Volonteri2008,Ferrara:2014wua,Pacucci:2015rwa,Madau:2014pta} under standard astrophysical scenarios. Recent James Webb Space Telescope (\textit{JWST}) observations have further complicated the picture by revealing a surprisingly large population of faint active galactic nuclei (AGN) at $z\gtrsim 4$, known as ``little red dots''~(LRDs; e.g. \cite{Kocevski2023,Greene2024,Matthee2024}). Most LRDs have significantly larger SMBH masses (given their host galaxy stellar masses) compared to local scaling relations~\citep{Pacucci2023}, signaling different pathways of BH and bulge mass growth \cite{Dayal:2024zwq}.

\begin{figure}
    \centering
    \includegraphics[width=0.98\linewidth]{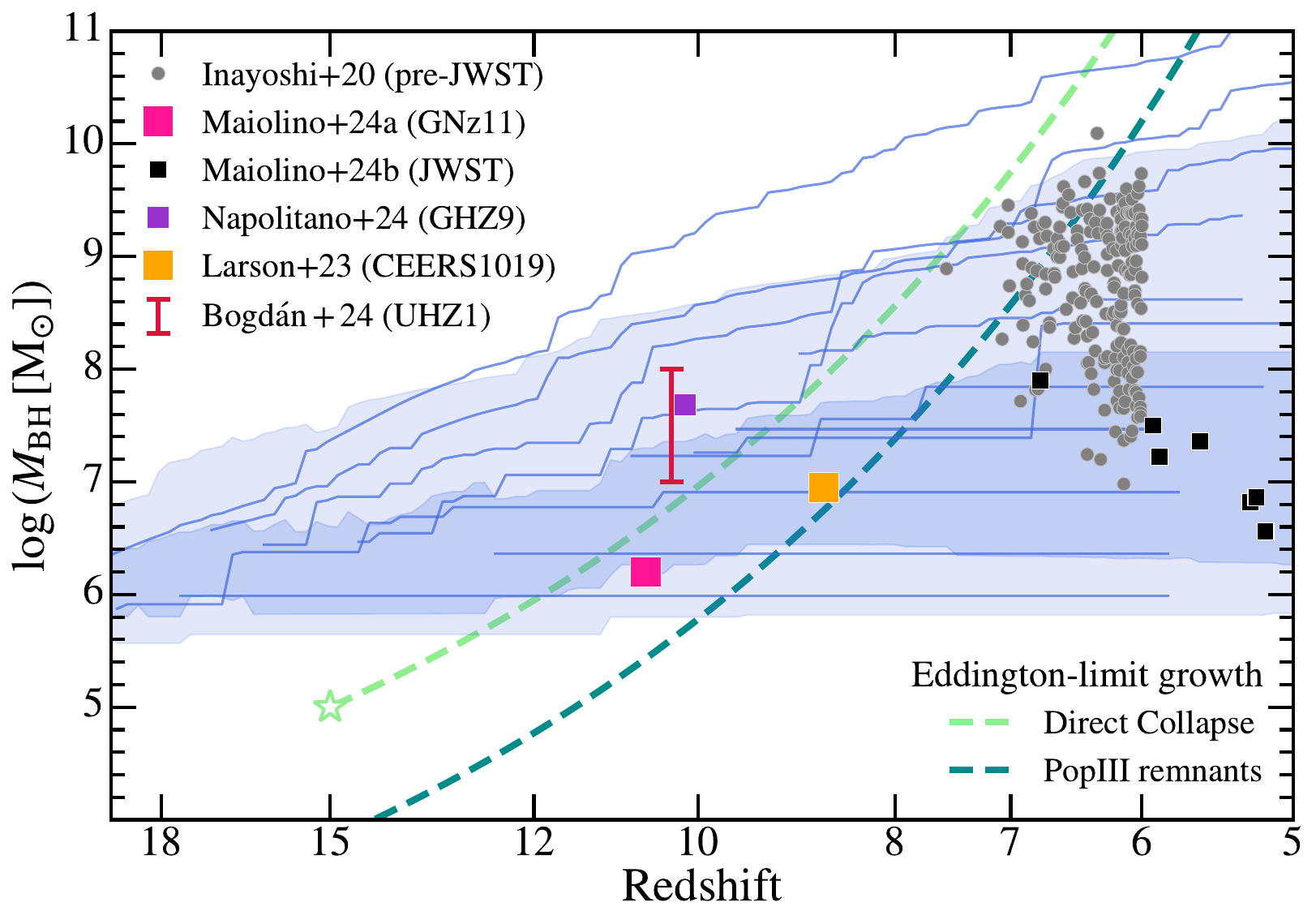}
    \caption{
        \textbf{An illustration of SMBH mass growth history.} The blue shaded regions show the 1 and 2-$\sigma$ scatters of the mass growth histories for dSIDM-seeded SMBHs (weighted by their Press-Schechter weights), with the faint blue lines showing selected mass growth histories. We compare it to the observed massive quasars at $z\gtrsim 6$~\citep{Inayoshi2020} and \textit{JWST}-identified AGN~\citep{Larson2023,Maiolino2024a,Maiolino2024b,Bogdan2024}. The two dashed lines represent Eddington-limit growth (with the Salpeter time $\sim 50\Myr$) from direct collapse BHs ($\sim 10^{5}\msun$ at $z\sim 15$) and PopIII stars ($\sim 300\msun$ at $z\sim 25$).
    }
    \label{fig:bh_growth_tracks}
\end{figure}

Massive SMBH seeding has been shown to be more feasible in alternative dark matter (DM) models featuring DM self-interactions~\citep[e.g.][]{Pollack:2014rja,Hu:2005cd,DAmico:2017lqj,Latif:2018kqv,Padilla:2020sjy,Feng:2021rst}. The ``gravothermal catastrophe'' of weakly-collisional self-gravitating systems can lead to run-away collapse at the center of DM halos~\citep{Lynden-Bell:1980xip,Balberg:2002ue,Koda:2011yb,Essig:2018pzq,Zhong:2023yzk,Tran:2024vxy} and the formation of compact objects. This phenomenon can be triggered in both elastic self-interacting DM (SIDM) models with heat conduction and dissipative SIDM (dSIDM) models with ``radiative'' cooling~\citep{Feng:2020kxv,Xiao2021,Lu:2024zwa,Buckley:2024eoe}. In particular, Ref.~\cite{Xiao2021} showed that dissipative SIDM can efficiently seed massive SMBHs at $z\gtrsim 6$ without violating any lower-redshift constraints. We focus on the same SIDM model that features hit-and-stick when two dark matter particles scatter with each other. This also represents a general class of dark matter with dissipative forces leading to a significant loss of momentum.

In this \textit{Letter}, we embed dSIDM gravothermal seeding in a merger-tree--based semi-analytic framework to predict the high-redshift SMBH population and its observational signatures. We show that this \emph{particle-physics alternative} to standard baryonic seeding scenarios can reproduce both the bright $z\gtrsim 6$ quasar population and the abundant faint AGN (``little red dots'') seen by JWST, including their elevated $M_{\rm BH}/M_*$ ratios. A central prediction is that luminous quasars represent only a subset of a much larger, predominantly nonactive SMBH population, which in turn enhances gravitational-wave signals: individual mergers can be detectable by LISA while the associated stochastic background remains consistent with current pulsar-timing limits. Although the required dissipative/self-interacting dark sector is exotic relative to collisionless $\Lambda$CDM, these correlated SMBH and GW signatures make the scenario sharply testable. Throughout this paper, $\log$ refers to the base-10 logarithm and $\ln$ to the natural logarithm.

\section{Methods}

\subsection{SMBH seeding from dSIDM}
\label{sec:seeding}

SIDM halos are subject to gravothermal collapse due to heat conduction and direct dissipative processes. Therefore, dissipative DM self-interactions can lead to the runaway collapse of the central region of a DM halo and ultimately form a massive BH seed~\citep{Xiao2021}. Although rarely investigated in astrophysical systems, such dissipative interactions are generic in hidden-sector extensions to the Standard Model, such as atomic DM~\citep[e.g.][]{Kaplan:2009de,Cyr-Racine:2012tfp}, exciting DM~\citep{Finkbeiner:2007kk,Finkbeiner:2014sja}, composite strongly interacting DM~\citep[e.g.][]{Fan:2013tia,Boddy:2014yra,Das:2017fyl}, and the quark nugget model~\citep{Gresham:2017cvl,Gresham:2017zqi}. Here, we consider a totally dissipative DM (all the kinetic energy in the center-of-momentum frame is lost during one DM-DM collision). In this scenario, the collapse/seeding timescale can be determined from halo parameters and the self-interacting cross-section~(\cite{Xiao2021,Spall})
\begin{align}
\label{eq:collapse_time}
 t_{\rm col} & \simeq 1\,{\rm Gyr}\,\left(\dfrac{c_{\rm halo}}{4}\right)^{-7/2} 
 \left[f(c_{\rm halo})\right]^{3/2} \left(\dfrac{\Delta_{\rm vir}}{200}\right)^{-7/6} \nonumber  \\
 & \left(\dfrac{\rho_{\rm crit}(z)}{\Omega_{\rm m}\rho_{\rm crit}(6)}\right)^{-7/6}
 \left(\dfrac{\sigma/m}{\rm 0.02\,cm^2/g}\right)^{-1}\left(\dfrac{M_{\rm vir}}{10^{13}M_{\odot}}\right)^{-1/3},
\end{align}
where $\sigma/m$ is the DM self-interaction cross-section (per unit mass), $M_{\rm vir}$ is the halo virial mass, $\Delta_{\rm vir}\equiv 200$ is the characteristic halo overdensity, $c_{\rm halo}$ is the halo concentration parameter, $f(x)\equiv \ln(1+x) - x/(1+x)$, and $\rho_{\rm crit}(z)$ is the critical density of the Universe at redshift $z$. BH seeding happens when the collapse time is shorter than the lifetime of the system, approximately given by
\begin{equation}
    \label{eq:collapse_condition_prac}
      t_{\rm col}(M_{\rm vir}(z), \sigma/m, z) \lesssim \epsilon \,t_{\rm H}(z),
\end{equation}
where $t_{\rm H}(z)\equiv 1/H(z)$ is the Hubble time and $\epsilon$ is an order-unity factor chosen to be $\pi/5$, motivated by the ratio of the halo dynamical time ($2\,\pi\,R_{\rm vir}/V_{\rm vir}$) to the Hubble time. Eq.~(\ref{eq:collapse_time}) suggests that rare DM halos with larger masses, higher concentrations, or at higher redshifts are more prone to seed massive BHs. We note the uncertainties in $\epsilon$ and should, in principle, be determined using cosmological simulations of dSIDM halos, which will be fulfilled in future works. Here, effectively $\epsilon\,(\sigma/m)$ determines the cosmological impact of this model, and our fiducial choice is $\epsilon\,(\sigma/m)=0.01 \cpm$, which corresponds to $\sigma/m \sim 0.016\cpm$. Such a low cross-section is completely unconstrained in local dwarf galaxies~\citep[e.g.][]{Shen2021,Shen2024}, and DM should behave as collisionless CDM in the Universe except for the massive rare halos at high redshifts considered here. 

\subsection{Semi-analytical model for SMBH evolution}
\label{sec:SAM}

We adopt a semi-analytical model (SAM) to predict the cosmological abundance of SMBHs formed via the direct collapse of SIDM halos. Here, we briefly overview the model~(see \cite{Spall} for details). The growth history of DM halos is traced through merger trees generated using \textsc{SatGen}~\footnote{https://github.com/shergreen/SatGen}~\cite{Jiang2021}, which is based on the Extended Press-Schechter formalism~\citep{Lacey1993} and the algorithm in Ref.~\cite{Parkinson2008,Jiang2014,Benson2017}. We uniformly sample in total $70$ root halos from $10^{8}$ to $10^{15} \msun$ at $z=4$ and trace their progenitors up to $z\simeq 20$. All the progenitors of one merger tree are weighted by the number density of the root halo at $z=4$. An SMBH seed is placed in a halo when the halo parameters meet the seeding criterion in Eq.~\ref{eq:collapse_condition_prac}. The initial mass of the seed is set as a constant fraction, $f_{\rm col}=3\times 10^{-3}$, of the instantaneous mass of the host halo, motivated by the simulation results in Ref.~\cite{Xiao2021}. Subsequently, as long as the host halo still satisfies the seeding criterion, we maintain the seed-to-host mass ratio as $f_{\rm col}$. The treatment relies on the assumption that, after the initial collapse of the DM halo, the accretion of DM onto the central SMBH seed will continue until a dynamical equilibrium between the SMBH seed and the host halo is reached. The accretion of baryonic matter for SMBHs~\citep{Volonteri2003,Volonteri2008,Natarajan2014} was found to be negligible compared to the large SMBH seed mass and DM accretion, even with the most aggressive parameter choices~\citep{Xiao2021}, which we ignore for the purpose of this work. During the merger of host halos, the dynamical friction against the DM background could drag the satellite halo and its SMBH towards the primary SMBH. We assume that this happens when the mass ratio of the two SMBH-plus-halo systems is larger than $0.3$~\citep{Volonteri:2002vz}. The dynamical friction time, $t_{\rm DF}$~(see \cite{Spall} for details), is the time delay of the SMBH merger to the halo merger. We effectively ignore the further hardening phase of the SMBH binary~\citep[e.g.][]{Begelman1980,Yu2002} through interactions with nuclear stars (and gas). The large viscosity induced by DM self-interactions can efficiently carry away angular momentum on time scales comparable to $t_{\rm col}$ and accelerate SMBH mergers compared to the standard baryon-driven scenarios~\citep{Feng:2020kxv,Xiao2021}.

\section{Results}

\subsection{Implications for the SMBH population at high redshifts}

Massive bright quasars observed at $z\gtrsim 6$ have long been viewed as strong constraints on SMBH seeding and growth in the early Universe~\citep[e.g.][]{Haiman2001,Inayoshi2020,Shen2026}. Canonical seeding mechanisms like PopIII star remnants~\citep[e.g.][]{Madau2001,Schneider2002,Volonteri2005} and run-away mergers in star clusters~\citep[e.g.][]{Begelman1978,Devecchi2009} will require a sustained super-Eddington accretion. Even for heavy seeds from directly collapsed pristine gas clouds~\citep[e.g.][]{Loeb1994,Eisenstein1995,Bromm2003,Lodato2006,Volonteri2008}, a substantial period of Eddington-limit growth is necessary. These are in tension with the small proximity zone sizes~\citep[e.g.][]{Eilers2021} and low inferred duty-cycles of bright quasars~\citep[e.g.][]{Eilers2024,Pizzati2024}, as well as the low Eddington ratios measured for some quasars~\citep[e.g.][]{Onoue2019,Juodzbalis2024}, although great uncertainties still exist in interpreting these results and pathways with episodic highly super-Eddington accretion cannot be ruled out. \textit{JWST} observations have pushed the redshift frontier of detecting massive SMBHs with the discovery of UHZ1~\citep{Bogdan2024}, CEERS1019~\citep{Larson2023}, GHZ9~\citep{Napolitano2024}, and GNz11 as an AGN~\citep{Maiolino2024a}. UHZ1 and GHZ9, in particular, would require super-Eddington accretion throughout the lifetime of a massive seed. The observed masses and redshifts of these sources are summarized in Fig.~\ref{fig:bh_growth_tracks}. They are compared to the growth track of dSIDM-seeded SMBHs selected from our merger tree, which can grow to $\gtrsim 10^{9}\msun$ at $z\sim 6$ without any baryonic accretion. The full dynamical range of massive pre-\textit{JWST} quasars and recently \textit{JWST}--identified AGN are well reproduced. The low-mass SMBHs ($M_{\rm BH} \sim 10^{6-8}\msun$) at $z\sim 6$ in this scenario stay in the host halo that satisfied the seeding criterion at earlier times but ceased accretion of DM later. Many properties of these ``left-over'' SMBHs appear consistent with the new class of low-luminosity AGN detected by \textit{JWST}, known as ``little red dots'' (LRDs; e.g. \cite{Matthee2024}) at $z\gtrsim 4$. The observed number density of LRDs is $\sim 10-100\times$ higher than the previously UV-selected AGN at similar UV magnitudes~\citep[e.g.][]{Kokorev2024}. 

\begin{figure}
    \centering
    \includegraphics[width=0.96\linewidth]{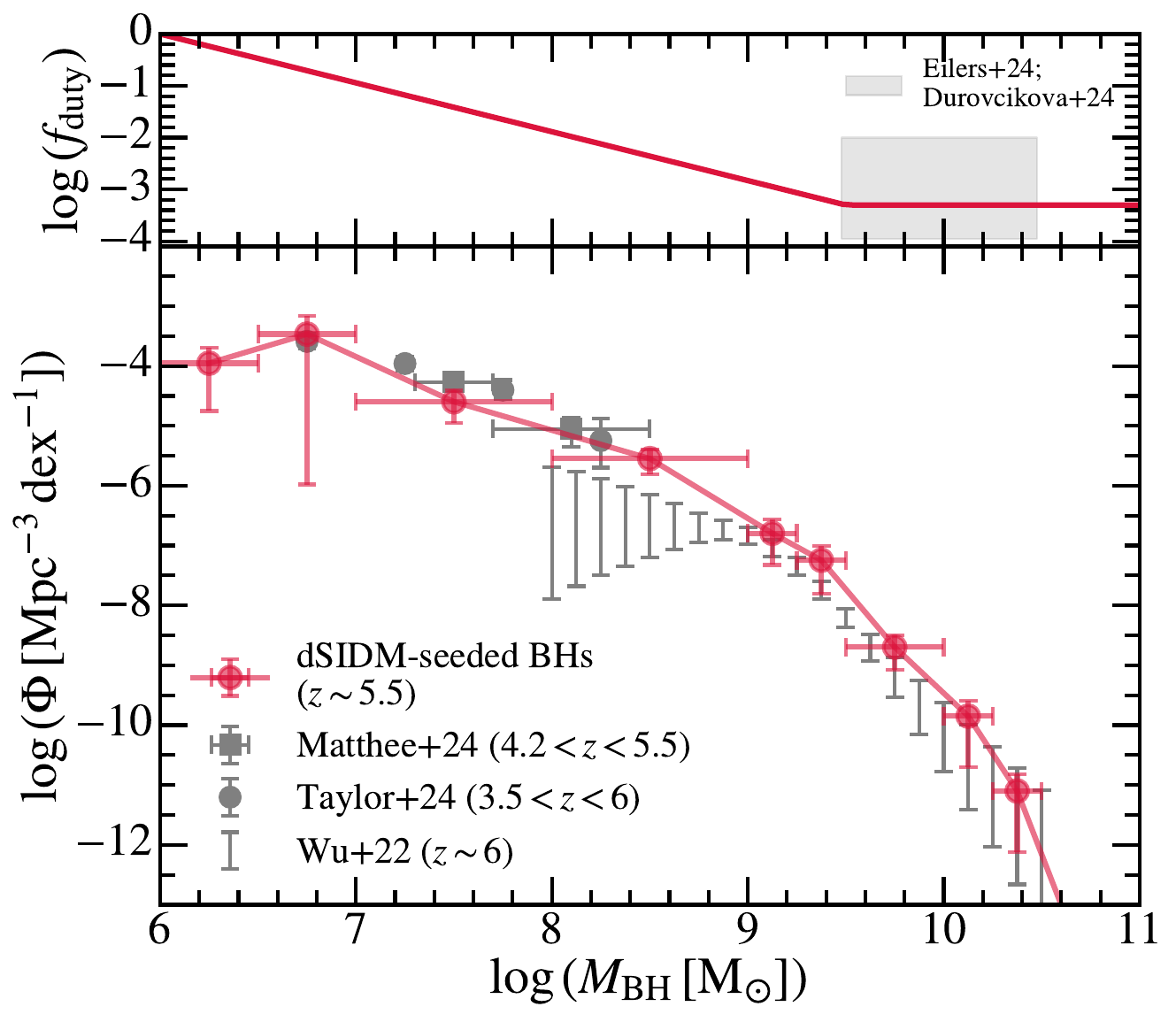}
    \caption{
        \textbf{SMBH mass function at $z \sim 5.5$.} We show the mass function of all dSIDM-seeded SMBHs in red, where a mass-dependent duty-cycle has been applied. We compare our results to the SMBH mass functions of observed broad-line AGN/quasars at similar redshifts~\citep{Wu2022,Matthee2024,Taylor2024}. In the top subpanel, we show the implied duty-cycle (tuned to match the observed SMBH mass function), which increases from $\lesssim 0.1\%$ for massive BHs to $100\%$ at $10^{6}\msun$. This implied $f_{\rm duty}$ for quasars agrees well with observational constraints from quasar clustering and proximity zone observations~\citep[e.g.][]{Eilers2024,Durovckova2024}.
    }
    \label{fig:smbhmf}
\end{figure}

In Fig.~\ref{fig:smbhmf}, we show the SMBH mass function in the dSIDM seeding scenario at $z\simeq 5.5$ and compare it to the observational constraints of broad-line AGN (BLAGN) at similar redshifts~\citep{Wu2022,Matthee2024,Taylor2024}. Since not all SMBHs are in the ``active'' phase, the AGN duty-cycle is a free parameter in this comparison. Through matching the SMBH mass function, we find empirically that $f_{\rm duty}\sim 5\times10^{-4}$ for quasars in the massive end. This value is consistent with recent observational constraints based on proximity zones~\citep[e.g.][]{Eilers2021} and quasar clustering~\citep[e.g.][]{Eilers2024,Pizzati2024,Durovckova2024}. The large population of dormant SMBHs in this scenario could leave strong GW signals from mergers, which will be investigated in the following section. Towards lower masses, an increased $f_{\rm duty}$ is implied and reaches $\sim 100\%$ at $\sim 10^{6}\msun$. However, we note that predictions for low-mass SMBHs are sensitive to the choice of $\epsilon$ and $\sigma/m$. For example, a higher value of $\epsilon,\sigma/m$ allows lower-mass black holes to remain coupled to the seeding mechanism, enhancing their abundance and thereby reducing the $f_{\rm duty}$ to match observational constraints. Meanwhile, for this low-mass tail of SMBHs, baryonic gas accretion could be relevant and is not yet modeled self-consistently here, which may also reduce the inferred $f_{\rm duty}$. Nevertheless, the predictions in the massive end are not affected by these uncertainties~(see Sec. IV of \cite{Spall} for a detailed discussion). 

\begin{figure}
    \centering
    \includegraphics[width=0.96\linewidth]{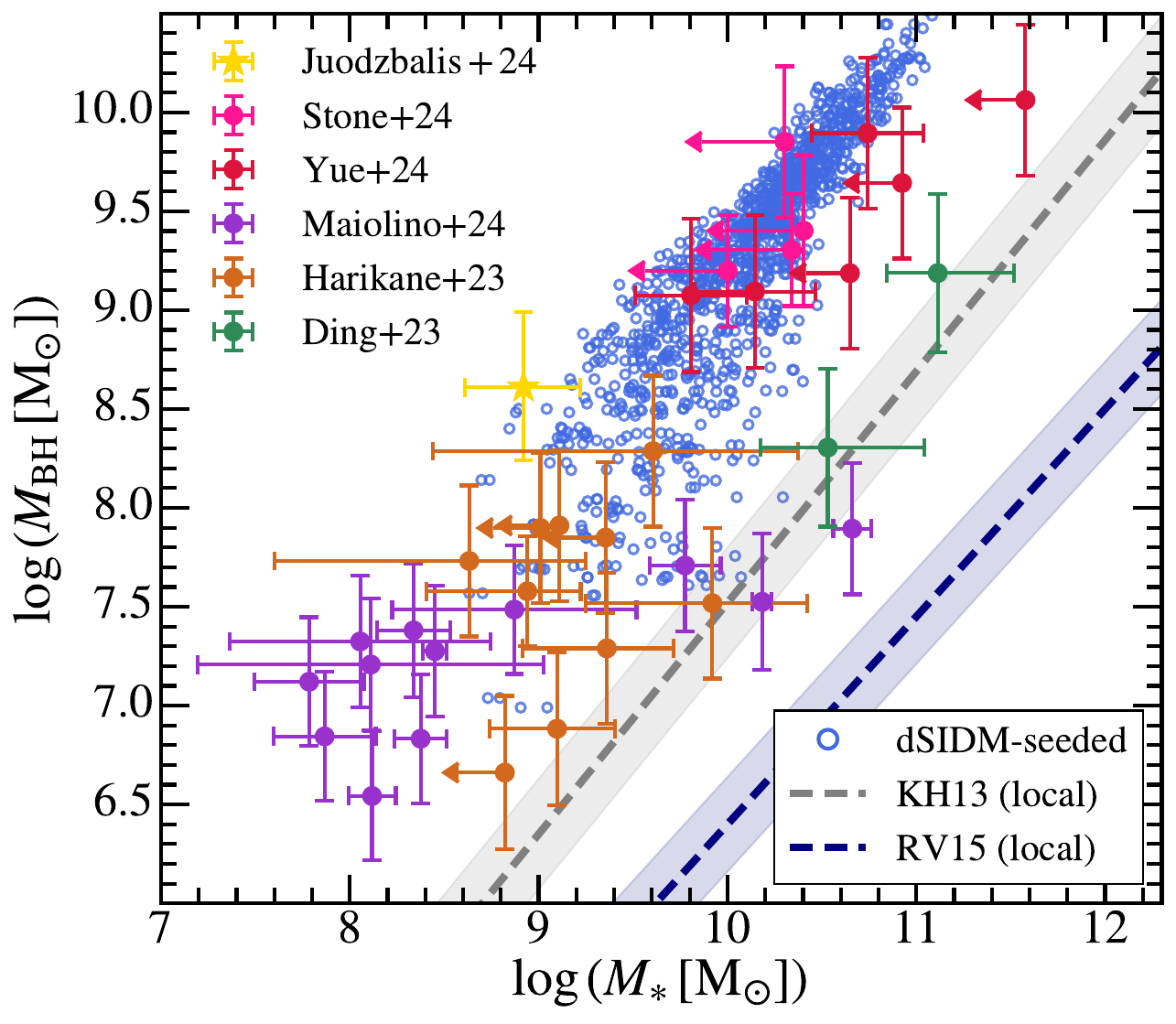}
    \caption{
        \textbf{SMBH Mass versus host galaxy stellar mass at $z \simeq 6$.} We show the dSIDM-seeded SMBHs in blue open circles. The galaxy stellar masses are computed from host halo mass using the UniverseMachine~\citep{Behroozi2019}. The model predictions are compared to observations of bright quasars~\citep{Ding2023,Stone2024,Yue2024}, LRDs~\citep{Harikane2023,Maiolino2024b}, as well as the local scaling relations~\citep{Kormendy2013,Reines2015} for reference.
    }
    \label{fig:smbh_vs_stellar_z6}
\end{figure}

Another puzzling fact of LRDs is that the host galaxy usually displays extremely compact morphology and has significantly lower stellar mass compared to the expectation from local scaling relations~\citep[e.g.][]{Pacucci2023}. In Fig.~\ref{fig:smbh_vs_stellar_z6}, we present the SMBH mass versus host galaxy stellar mass relation derived from our SAM at $z\simeq 6$. We utilized the observationally constrained stellar-to-halo mass ratios~\citep{Behroozi2019} to compute galaxy stellar mass, with an additional log-normal scatter of $0.1$ dex to account for the uncertainties in the scaling relation. The dSIDM-seeded SMBH masses have a tight correlation with galaxy stellar mass with a ratio of $\sim 0.1$, inherited from the correlation with host halo mass as discussed in Sec.~\ref{sec:seeding}. At $M_{\ast}\lesssim 10^{10}\msun$, SMBHs start to decouple from our seeding mechanism, and the passive evolution through halo mergers decreases the SMBH-to-stellar-mass ratio to $\sim 0.01 - 0.1$. These values are in surprisingly good agreement with the overly massive LRDs~\footnote{The BH mass in some LRDs could be overestimated due to the uncertainties in dust extinction (versus the Balmer break due to extremely dense gas, \cite{Naidu2025}), broadening of Balmer lines by dense ionized gas cacoon~\citep[e.g.][]{Naidu2025,Rusakov2025}, or breakdown of the calibrations of BH mass indicators from low-redshift AGN.} found by \textit{JWST}. For comparison, we show the LRDs in Ref.~\cite{Harikane2023, Maiolino2024b,Juodzbalis2024} and the bright quasars from Ref.~\cite{Ding2023,Stone2024,Yue2024}. The local $M_{\rm BH}$–$M_{\rm bulge}$ relation from Ref.~\cite{Kormendy2013} and $M_{\rm BH}$–$M_{\ast}$ relation from Ref.~\cite{Reines2015} are shown for reference. 

\begin{figure}
    \centering
    \includegraphics[width=1.05\linewidth]{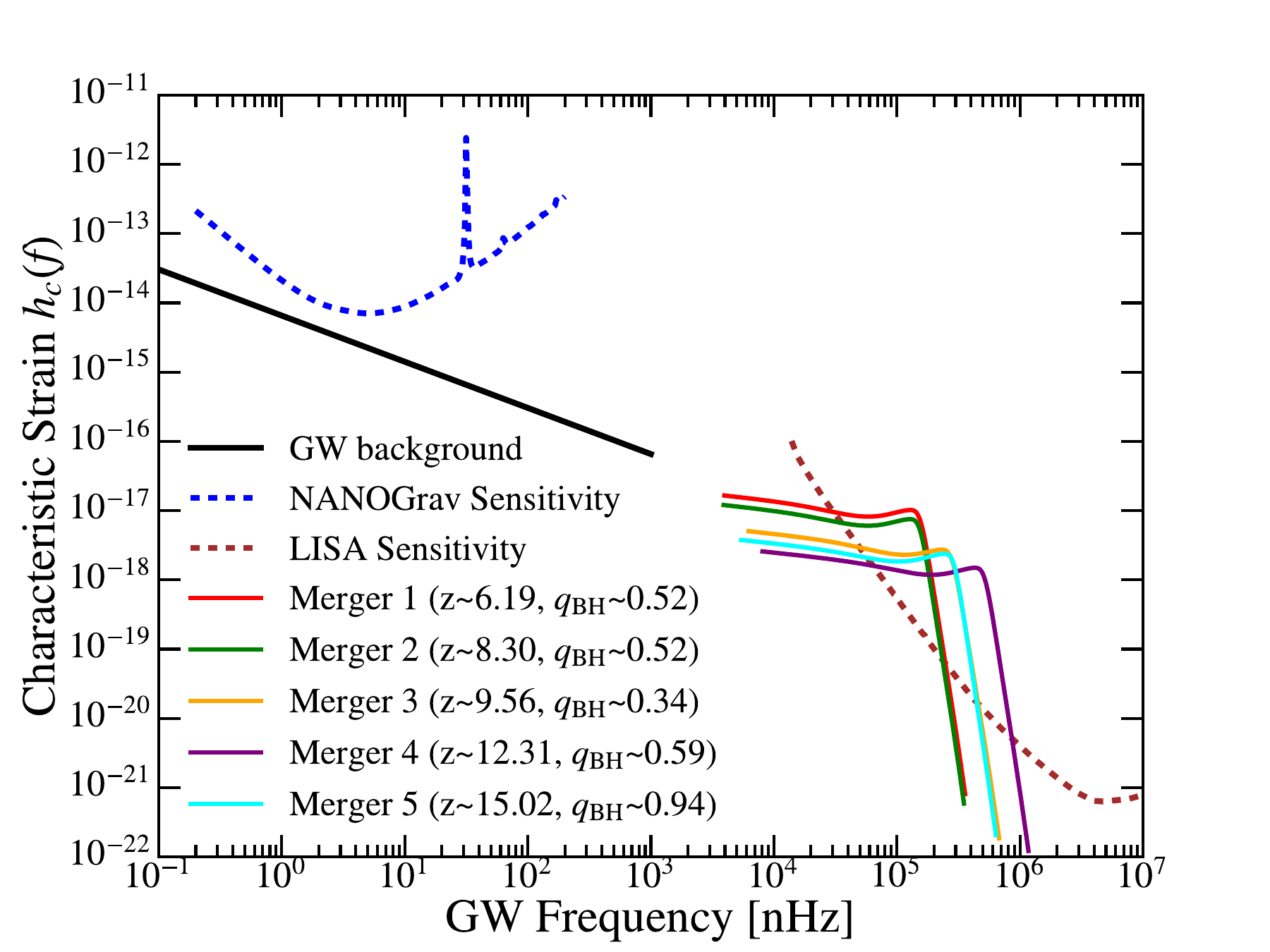}
    \caption{
        \textbf{GW spectrum from SMBH mergers.} We show both the stochastic background generated by all mergers of dSIDM-seeded SMBHs and signals from selected individual SMBH merger events. These are compared to the sensitivity curves of NANOGrav and LISA. The GW background contributed by this population remains below the NANOGrav sensitivity and, thus, will not violate any current constraints. Meanwhile, the selected merger events are observable with LISA. Here, $q_{\rm BH}\equiv M^2_{\rm BH}/M^1_{\rm BH}\leq 1$ denotes the mass ratio of the two merging SMBHs.
        }
    \label{fig:gwspec}
\end{figure}

Conceptually, the dSIDM seeding (or dark sector seeding in general) scenario breaks the connection between SMBH mass growth and galaxy bulge build-up. Mass growth does not rely on the fueling of baryonic gas to the vicinity of the SMBH, which should inevitably fragment and trigger star-formation in the proto-bulge along the way. The gas reservoir around dSIDM-seeded SMBHs can be expelled by processes like AGN feedback, leaving little star-formation in the host galaxy and the low duty-cycle in the quasar phase. However, low-mass SMBHs can still spend most of their lifetime in the active phase, as implied by the comparison of SMBH mass functions above. They appear as the observed low-luminosity AGN (LRDs), although baryonic accretion is not responsible for their past mass growth. These low-mass SMBHs can preserve dense gas structures around them dragged gravitationally by the inflowing dSIDM, which give rise to the Balmer absorption lines and Balmer breaks seen in LRDs~\citep[e.g.][]{Inayoshi2025,Naidu2025} and resulting in X-ray weakness~\citep[e.g.][]{Yue2024-xray,Maiolino2024c}. Furthermore, as shown in Ref.~\cite{Xiao2021}, the redshift dependence in Eq.~\ref{eq:collapse_time} implies that dSIDM-seeded SMBHs will only appear in a redshift window that agrees with the redshift distributions of LRDs. For example, the cumulative number of halos where Eq.~\ref{eq:collapse_condition_prac} is satisfied peaks around $z\sim 9$ and declines by an order-of-magnitude towards $z\sim 4$, assuming the same survey field-of-view and depth.

\begin{figure}
    \centering
    \includegraphics[width=\linewidth]{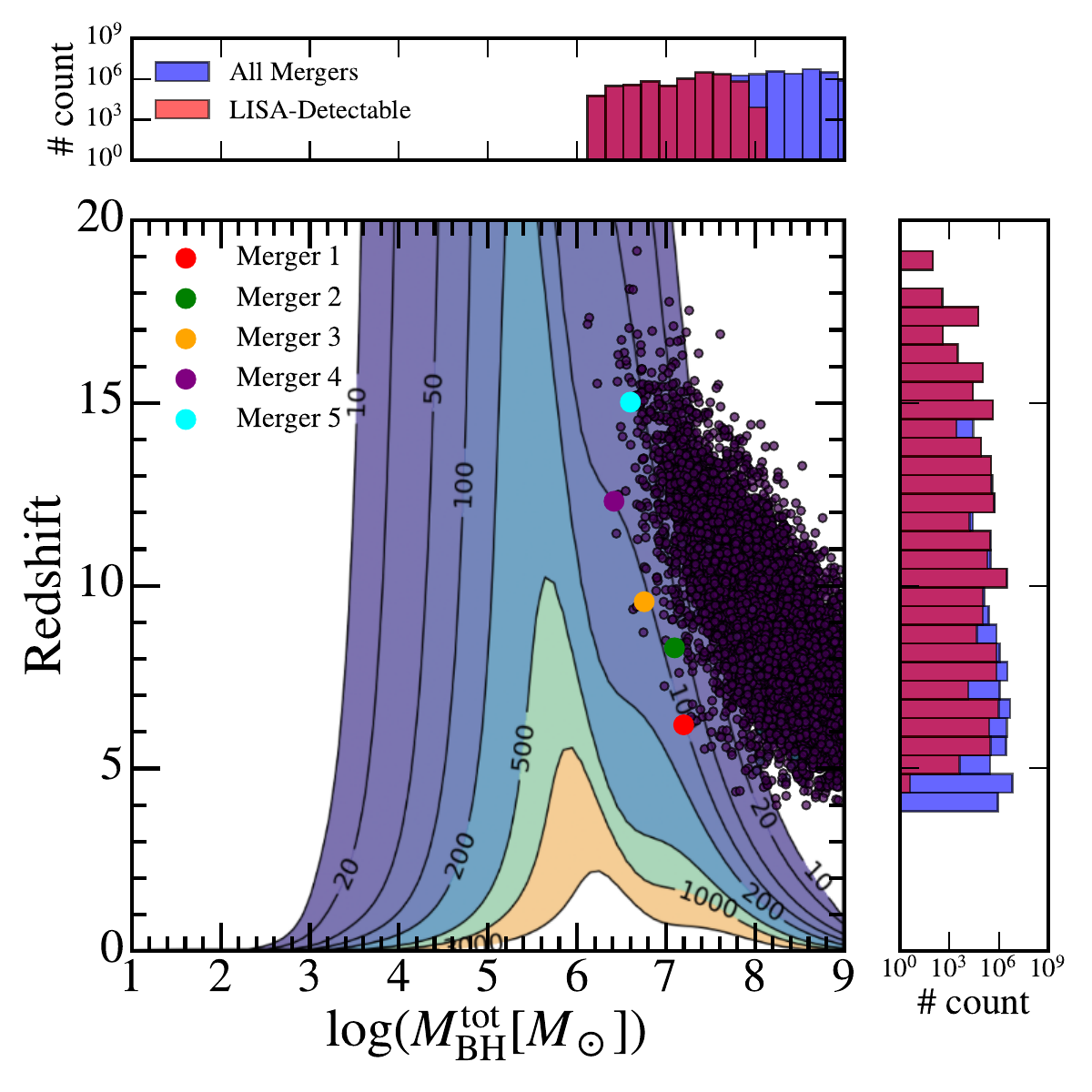}
    \caption{
    \textbf{Redshift versus total mass ($M^{\rm tot}_{\rm BH}=M^1_{\rm BH} + M^2_{\rm BH}$) of SMBH merger events, compared to the LISA sensitivity contour.} Different regions are color-coded by the SNRs of LISA observations, computed assuming a fixed mass ratio $q_{\rm BH}=0.25$ of merging SMBHs. The histograms in the top and right subpanel show the mass and redshift distributions (weighted by the differential comoving volume at the corresponding redshift) of all SMBH merger events and those of LISA-detectable events.
    }
\label{fig:smbh_mergers_lisa_panels}
\end{figure}

\subsection{Gravitational wave signals}

Two types of GW signals are expected from this large population of massive SMBHs at high redshift. The first is the GW background from unresolved SMBH mergers. We compute it based on the SMBH merger events recorded in our SAM. In Fig.~\ref{fig:gwspec}, we show the characteristic strain of the GW background from dSIDM-seeded SMBHs as a function of frequency and compare it to the sensitivity curves of NANOGrav and LISA. The GW background contributed by this population remains consistent with the current NANOGrav constraints~\citep{NANOGrav:2023ctt}. The second type of signal is GWs from individual SMBH merger events. In Fig.~\ref{fig:smbh_mergers_lisa_panels}, we illustrate the detectability of these events using LISA, overlaying distributions of SMBH merger events with the signal-to-noise ratio (SNR) contours of LISA. The computation of the GW spectrum of SMBH mergers and the LISA SNRs is described in \cite{Spall}. The spectra of five selected events are also shown in Fig.~\ref{fig:gwspec}. LISA can detect the low-mass tail of the dSIDM-seeded SMBHs with SNRs up to $\sim 100$. The total SMBH mass distribution of detectable events peaks around $M^{1}_{\rm BH}+M^{2}_{\rm BH} \sim 10^{7}\msun$ and the redshift distribution peaks around $z\sim 12$. The total number of events will be significantly higher than in scenarios with lower-mass seeds, as the rapid BH growth required to reach bright quasar masses implies a much lower abundance of dormant SMBHs. The LISA predictions presented in this section depend on the choice of $\sigma/m$, since a lower $\sigma/m$ will restrict dSIDM seeding in more massive haloes and the resulting binary BH mass can be shifted to outside LISA's most sensitive window. Therefore, our fiducial choice lies in a sweet spot of producing detectable GW signals for LISA while not overriding local Universe constraints on dSIDM~\citep{Spall}.

\section{Conclusions}
We have studied the SMBH population at $z\gtrsim 4$ in a dSIDM gravothermal-seeding scenario using a semi-analytic merger-tree approach, deriving population-level and multi-messenger consequences. \emph{Key results are:} (i) massive bright quasars and low-mass faint AGN (LRDs) arise naturally, with $M_{\rm BH}/M_*$ consistent with JWST observations; (ii) because SMBH masses are largely set at collapse rather than by baryonic accretion and bulge growth, the predicted evolution differs qualitatively from the low-redshift, baryon-regulated SMBH--galaxy coevolution picture; and (iii) the large implied SMBH number density requires a sub percent-level duty cycle for bright quasars (rising toward unity for low-mass AGN), predicting a substantial nonactive SMBH population whose mergers yield GW events detectable by LISA while remaining compatible with current NANOGrav constraints on the stochastic background. More broadly, similar seeding channels can arise in other SIDM realizations, and forthcoming JWST, LISA, and pulsar-timing data will provide decisive tests of DM-driven SMBH formation.

\section*{Acknowledgements}
We thank Akshay Ghalsasi, David Dunsky, Josh Foster, Weixiang Feng, Oliver Zier, Yueying Ni, Rohan Naidu, and Philip Hopkins for their useful comments and discussions. HX is supported by Fermi Forward Discovery Group, LLC under Contract No. 89243024CSC000002 with the U.S. Department of Energy, Office of Science, Office of High Energy Physics. FJ is supported by the National Science Foundation of China grant No. 12473007.

\bibliographystyle{apsrev4-2}
\bibliography{ref}

\clearpage
\onecolumngrid
\begin{center}
\textbf{\large Supplemental Materials}
\end{center}
\hspace{5mm}
\twocolumngrid
\setcounter{equation}{0}
\setcounter{figure}{0}
\setcounter{table}{0}
\setcounter{section}{0}
\setcounter{page}{1}

\section{Semi-analytical model for SMBH seeding and evolutions}
\label{appsec:sam}

\subsection{Dissipation time scale}

For SMBH formation driven by dSIDM, we can quantitatively determine the collapse timescale in DM halos given its mass and concentration. First, we consider virialized DM halos with the Navarro–Frenk–White (NFW, \cite{Navarro1996}) density profiles
\begin{equation}
    \rho(r) = \dfrac{\rho_{\rm s}}{\frac{r}{r_{\rm s}}\left(1+\frac{r}{r_{\rm s}}\right)^2},
    \label{eq:nfw}
\end{equation}
where $\rho_{\rm s}$ and $r_{\rm s}$ are the scale density and radius of the halo. The virial mass of the halo is
\begin{equation}
     M_{\rm vir}=\frac{4}{3}\pi\, c_{\rm halo}^3\,r_{\rm s}^3\, \Delta_{\rm vir}\, \rho_{\rm crit}(z),
\end{equation}
where $c_{\rm halo} = \frac{r_{\rm vir}}{r_{\rm s}}$ is the halo concentration, $\rho_{\rm crit}(z)$ is the critical density of the Universe at redshift $z$, and $\Delta_{\rm vir}=200$ is the overdensity of the halo with respect to the critical density. From these relations, one finds the scale density of DM halos
\begin{equation}\label{eq:determine_rho0}
    \rho_{\rm s}=\frac{M_{\rm vir}}{4\pi r_{\rm s}^3(z) f(c_{\rm halo})}=\frac{c_{\rm halo}^3}{3\,f(c_{\rm halo})}\,\Delta_{\rm vir}\,\rho_{\rm crit}(z),
\end{equation}
where $f(x) \equiv \ln{(1+x)} - x/(1+x)$. 
DM particles can lose their kinetic energy through dissipative self-interactions. Here, we assume that such interactions (collisions) are isotropic and have a velocity-independent cross-section. For simplicity, the fraction of kinetic energy loss in the center-of-momentum frame per collision is assumed to be unity (i.e. a hit-and-stick model). This energy loss fraction in general particle physics models is a free parameter, but for the phenomenology of BH seeding considered here, it is degenerate with the self-interaction cross-section $\sigma/m$.

The average timescale for a particle to encounter one such collision can be estimated as
\begin{equation}
    \label{eq:relaxation_time}
    t_{\rm coll} = \dfrac{1}{\alpha\,\rho\,\sigma_{\rm v}\,(\sigma/m)},
\end{equation}
where $\sigma_{\rm v}$ is the DM one-dimensional velocity dispersion, $\alpha=\sqrt{16/\pi}$ is a constant factor assuming hard-sphere-like scattering and a Maxwell-Boltzmann velocity distribution, and $\sigma/m$ is the self-interacting cross-section per particle mass. If the velocity field in a DM halo is isotropic, $\sigma_{\rm v}(r)$ can be obtained by solving the spherical Jeans equation~\citep{Lokas2001}
\begin{align}
    & \sigma_{\rm v}(r) = \sqrt{4\pi G\, \rho_{\rm s}\, r_{\rm s}^2\, F(r/r_{\rm s})} ,\nonumber \\
    & F(x) \equiv \dfrac{1}{2} x (1+x)^{2} \Big[ \pi^2 - \ln{(x)} - \dfrac{1}{x} \nonumber \\
    & - \dfrac{1}{(1+x)^2} - \dfrac{6}{1+x} + \Big( 1 + \dfrac{1}{ x^2} - \dfrac{4}{x} - \dfrac{2}{1+x}\Big) \nonumber \\ 
    & \times \ln{(1+x)} + 3 \ln^{2}{(1+x)} - 2{\rm Li}_{2}(-x) \Big], 
    \label{eq:velocity-dispersion}
\end{align}
where ${\rm Li}_{2}(x)$ is the dilogarithm. The timescale for local kinetic energy to dissipate through such collisions is
\begin{align}
    t_{\rm diss}(r) = \dfrac{1}{\beta\, \rho(r)\,\sigma_{\rm v}(r)\,(\sigma/m) },
    \label{eq:dissipation_timescale}
\end{align}
where the constant fudge factor $\beta=4\alpha/3$~\citep{Shen2021}, assuming again a Maxwell-Boltzmann velocity distribution. 

Rapid kinetic (thermal) energy dissipation will inevitably result in the gravitational collapse
of the central halo. The collapse timescale should be on the same order as the dissipation
time, $t_{\rm col} = A\,t_{\rm diss}$, where the order-unity factor $A=1.06$ is determined by simulations of isolated
DM halos in Ref.~\cite{Xiao2021}. In these simulations, the collapsed mass fraction of the DM halo was found to be $f_{\rm col} \simeq 3\times 10^{-3}$, which corresponds to the collapse radius (the radius where DM particles in the initial mass distribution will fall into the halo center and collapse) of $\sim 0.07\,r_{\rm s}$. Therefore, we evaluate $t_{\rm col}$ at this radius and obtain Eq.~1 in the main text.

\subsection{Halo merger trees}

We trace SMBHs formed from the dSIDM scenario above using halo merger trees. These merger trees are generated using the \textsc{SatGen}~\footnote{https://github.com/JiangFangzhou/SatGen} code~\cite{Jiang2021}, which is based on the Extended Press-Schechter (EPS) formalism~\cite{Lacey1993} and the algorithm introduced in \cite{Parkinson2008,Jiang2014,Benson2017}. When generating the merger trees, we uniformly sample $10$ halos per dex of halo mass ranging from $10^{8}$ to $10^{15} \msun$ at $z=4$ and trace their progenitors up to $z\simeq 20$, with an adaptive halo mass resolution that is $5$ dex lower than the mass of the root descendant halo at $z=4$. The merger tree traces the mass and concentration of each halo from the time it enters the tree (becomes more massive than the mass resolution of the tree) to the time when it merges into a more massive halo. The halo concentration is obtained from an empirical relation calibrated via simulations~\citep{Zhao2009}, which relates the main branch (the branch that tracks the most massive progenitor) merging history to the concentration parameter by
\begin{equation}
    c_{\rm halo}(M_{\rm vir}, z) = \left[4^8 + \left(t(z)/t_{\rm 0.04}(M_{\rm vir}, z)\right)^{8.4} \right]^{1/8},
    \label{eq:concentration_Zhao09}
\end{equation}
where $t(z)$ is the cosmic time at redshift $z$ and $t_{\rm 0.04}$ is the cosmic time when the host halo has assembled $4\%$ of its instantaneous mass, $M_{\rm vir}(z)$. In principle, the gravitational impact of baryonic matter ({\em e.g.} adiabatic contraction of DM, \cite{Blumenthal1986,Ryden1987}), star formation, and subsequent feedback processes could potentially affect the structure of high redshift halos. However, self-consistently modeling the baryonic content of high-redshift galaxies is beyond the scope of this paper, and we defer a detailed analysis of this aspect to follow-up work. All the progenitors of one merger tree are weighted by the number density of the final halo sampled at $z=4$, determined analytically by the halo mass function from the \textsc{hmf} code~\citep{Murray2013}. We adopt the following cosmological parameters for both the halo mass function and merger tree calculations: $H_0 = 67.66 \, \mathrm{km/s/Mpc}$, $\Omega_{\rm m} = 0.3111$, $\Omega_{\rm b} = 0.049$, $n_{\rm s} = 0.9665$, and $\sigma_8 = 0.8102$ from the Planck 2018 TT,TE,EE+lowE+lensing+BAO constraints~\citep{Planck2018}. Halos are defined based on the spherical overdensity with respect to $\rho_{\rm crit}(z)$ with a top-hat window function in real space and $\Delta_{\rm vir}=200$. We adopt the transfer function in Ref.~\cite{Eisenstein1998} and the halo mass function fitting function in Ref.~\cite{Tinker2010}. The halo mass function at an arbitrary redshift can be computed as
\begin{equation}
    \Phi(M_{\rm BH},z) = \sum_{\rm i} \dfrac{W(M^{\rm i}_{\rm vir}, z_0)}{n_{\rm sample}}\,\dfrac{{\rm dN_{\rm i}(z)}}{{\rm d}\log{M_{\rm BH}}},
    \label{eq:hmf-tree}
\end{equation}
where $n_{\rm sample}=10$ is the number of halos sampled per dex of halo mass, $W(M^{\rm i}_{\rm vir}, z_0)$ is the halo mass function of the root halo of the ith merger tree at the redshift of sampling $z_{\rm 0}=4$ (also referred to as the Press-Schcheter weight), and ${\rm dN_{\rm i}(z)}/{\rm d}\log{M_{\rm BH}}$ is the number of SMBHs in the $\log{M_{\rm BH}}$ bin at the target redshift $z$ in the ith merger tree. Other statistical quantities can be calculated with the same Press-Schcheter weight.

\subsection{SMBH seeding, growth, and mergers} 

An SMBH seed is initialized when the halo meets the seeding criterion in Eq.~2 in the main text. The initial mass of the seed is set as a constant fraction, $f_{\rm col}=3\times 10^{-3}$, of the instantaneous mass of the host halo. Subsequently, as long as the host halo still meets the seeding criterion, we maintain the seed-to-host mass ratio as $f_{\rm col}$. The treatment relies on the assumption that, after the initial collapse of the DM halo, the accretion of DM onto the central SMBH seed will continue until a dynamical equilibrium between the SMBH seed and the host halo is reached. This is motivated by the universal $f_{\rm col}$ found in idealized simulations~\citep{Xiao2021}, regardless of DM halo mass, redshifts, spin, and DM self-interaction cross-sections. Therefore, the growth history of a SMBH is tightly correlated with that of its host halo until the seeding criterion is no longer satisfied. The subsequent growth of SMBHs they host will no longer be affected by DM physics but by hierarchical mergers of SMBHs during halo mergers and accretion of baryonic matter. In Ref.~\cite{Xiao2021}, we have experienced the ``merger-driven'' accretion of baryonic matter for SMBHs~\citep[e.g.][]{Volonteri2003,Volonteri2008,Natarajan2014} and found that it is negligible compared to the large SMBH seed mass and DM accretion even with the most aggressive parameter choices that would violate the local $M_{\rm BH}-\sigma_{\ast}$ relation. 

During the merger of host halos, the dynamical friction against the DM background could drag the satellite SMBH towards the primary SMBH and a bound SMBH binary will form. We assume that this happens when the mass ratio of the two SMBH-plus-halo systems is larger than $0.3$, as suggested in Ref.~\cite{Volonteri:2002vz}. The time delay between the formation of a bound SMBH binary and the original halo merger is assumed to be the dynamical friction time $t_{\rm DF}$ of the satellite halo during each halo merger event~\citep{Lacey1993,Binney1987}
\begin{equation}
    t_{\rm DF} = 0.495\,\dfrac{1+\xi}{\xi}\,\dfrac{1}{H(z)\,\sqrt{\Delta_{\rm vir}}\,\ln{(1+\xi)}},
\end{equation}
where $\xi$ is the mass ratio of the two merging halo-plus-SMBH systems. We note that several effects are not considered in this simple treatment. First, we do not model the subsequent hardening phase of the binary~\citep{Begelman1980,Yu2002} and treat the bound binary as a single SMBH right after the merger. In the classical picture, the hardening is driven by the three-body interactions of SMBH binary with nuclear stars~\citep[e.g.][]{Quinlan1996} and gas~\citep[e.g.][]{Haiman2009}. However, the accretion of SMBHs considered here is dominated by dSIDM with a large viscosity generated by DM self-interactions~\citep{Feng:2020kxv,Xiao2021}, which can efficiently dissipate angular momentum and accelerate the merger. Secondly, we do not model the complicated interactions of SMBH triplets, which could form through hierarchical mergers. The intruding SMBH can facilitate the coalescence of the binary through close three-body interactions and Kozai-Lidov oscillations~\citep[e.g.][]{Iwasawa2006,Hoffman2007}. Meanwhile, the lightest SMBHs can be ejected through the interactions, and the recoil due to the GW emission after the binary merger can also lead to the ejection of the remnant SMBH~\citep[e.g.][]{Fitchett:1983qzq}. These processes could introduce order-unity uncertainties to the SMBH occupation fractions and masses in the regime when the seeding criterion is no longer met, but they should not change the conclusion about the SMBH population that is tightly coupled to the dSIDM seeding mechanism. 

\begin{figure*}
    \centering
    \includegraphics[width=\linewidth]{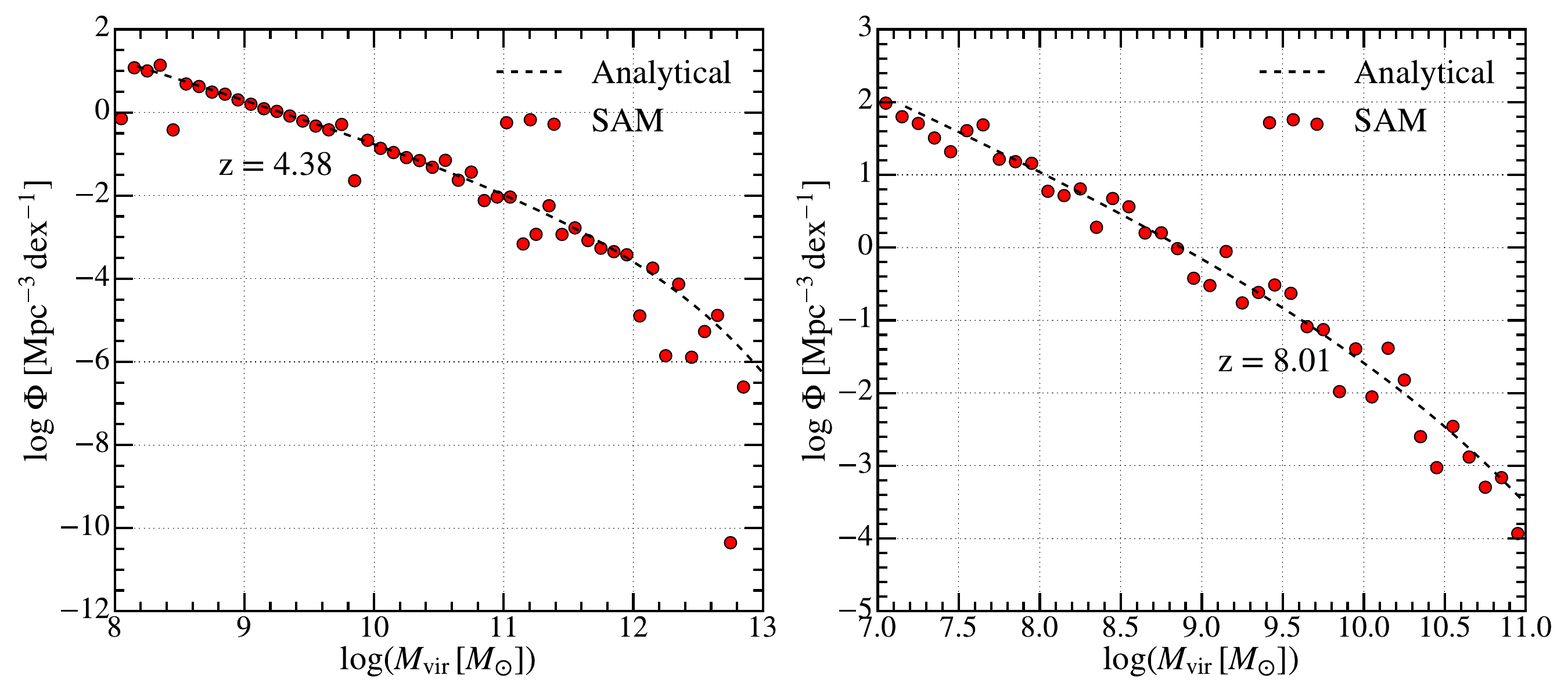}
    \caption{
        \textbf{Halo mass function at redshift $z = 4.38$ and $z = 8.01$}. 
        We compare the analytical halo mass functions (dashed black lines) and the halo mass functions derived from our SAM (merger trees) using the weight scheme in Eq.~\ref{eq:hmf-tree} (red points). The SAM results closely follow the analytical halo mass functions, demonstrating the effectiveness of the weighting approach. 
    }
    \label{fig:hmf_comparison}
\end{figure*}

\section{Gravitational wave spectrum}
\label{appsec:gw}

\subsection{Stochastic GW spectrum}

We compute the stochastic GW signal in the PTA band of our simulations with the formalism in Ref.~\cite{Phinney:2001di} and Ref.~\cite{Sesana2008}. We assume binaries to be on quasi-circular orbits and the chirp mass of a binary is defined as $\mathcal{M}_{\rm c} \equiv \frac{(M_{1}^{\rm BH} M_{2}^{\rm BH})^{3/5}}{(M_{1}^{\rm BH} + M_{2}^{\rm BH})^{1/5}}$. The characteristic GW amplitude at frequency $f$ in the observer's frame is~\citep{Phinney:2001di,Toubiana2024}
\begin{equation}
    h^2_{\rm c}(f) = \dfrac{4\,G^{5/3}}{3\,\pi^{1/3}\,c^2} \, f^{-4/3}\, \sum_{\rm i} \dfrac{W(M^{\rm i}_{\rm vir}, z_0)}{n_{\rm sample}} \, \sum_{\rm j} \dfrac{\mathcal{M}^{5/3}_{\rm c,i,j}}{(1+z_{\rm i,j})^{1/3}}
\end{equation}
The GW spectrum from SMBH mergers was derived using halo merger trees generated with \texttt{SatGen}. The NANOGrav PTA is sensitive to the stochastic GW background in roughly the frequency range $1 - 100$ nHz, and is ideal for probing SMBH mergers with large chirp masses at high redshifts. We obtain the stochastic GW background sensitivity curve with the NANOGrav 15-year data set in Ref.~\cite{NANOGrav:2023ctt}, including the contributions from white noise, red noise, and GW background self-noise.

\subsection{GW spectrum of individual events}

The characteristic strain $h_c(f)$ used to compute the signal-to-noise ratio (SNR) was obtained from waveform models using the PhenomD model~\citep{Husa2015, Khan2016}. The waveforms were generated with individual component spins set to 0.8 and 0.6 (reference values set to reproduce the SNRs shown in the LISA science paper \cite{Amaro-Seoane2017}), and waveform evolution was simulated starting from one year before the merger, which corresponds to the evolution of the binary's gravitational waveform from an early inspiral phase to the final merger and ringdown stages. The frequency evolution of the signal was obtained over a range of $10^{-5}$ Hz to $1$ Hz. The SNR is then computed as
\begin{equation}
    {\rm SNR}^2 = \frac{16}{5} \int \frac{h_{\rm c}^2(f)}{f^2 (S_{\rm n}(f)+S_{\rm c}(f))} \, {\rm d}f,
\end{equation}
where $h_{\rm c}(f)$ was interpolated from the PhenomD waveform data. The 16/5 prefactor accounts for LISA's sky-averaged response to GWs, which comes from averaging over all sky locations, inclinations, and polarizations. 

Following Ref.~\cite{Robson2018}, we model the LISA noise spectrum as
\begin{equation}
    S_{\rm n}(f) = \frac{10}{3 L^2} \left[P_{\rm oms}(f) + \frac{4 P_{\rm acc}(f)}{(2 \pi f)^4}\right]\left(1 + \frac{6}{10} \frac{f^2}{f_\ast^2}\right),
\end{equation}
where $L = 2.5$ Gm is the LISA arm length, and $f_\ast = 19.09$ mHz is the transfer frequency. The two instrumental noise components are defined as
\begin{align}
    & P_{\rm oms}(f) = (1.5 \times 10^{-11})^2 \left(1 + \left(\frac{2 \times 10^{-3}}{f}\right)^4\right), \nonumber \\
    & P_{\rm acc}(f) = (3 \times 10^{-15})^2 \left(1 + \left(\frac{0.4 \times 10^{-3}}{f}\right)^2\right) \nonumber \\
    &\times\left(1 + \left(\frac{f}{8 \times 10^{-3}}\right)^4\right),
\end{align}
and the noise from galactic binaries is modeled as
\begin{equation}
    S_{\rm c}(f) = A\, f^{-7/3} e^{-f^\alpha + \beta f \sin(\kappa f)} \left[1 + \tanh\left(\gamma(f_{\rm k} - f)\right)\right],
\end{equation}
where $A = 9 \times 10^{-45}$, $\alpha = 0.138$, $\beta = -221$, $\kappa = 521$, $\gamma = 1680$, and $f_{\rm k} = 1.13 \times 10^{-3}$ Hz.

\begin{figure}[h!]
    \centering
    \includegraphics[width=\linewidth]{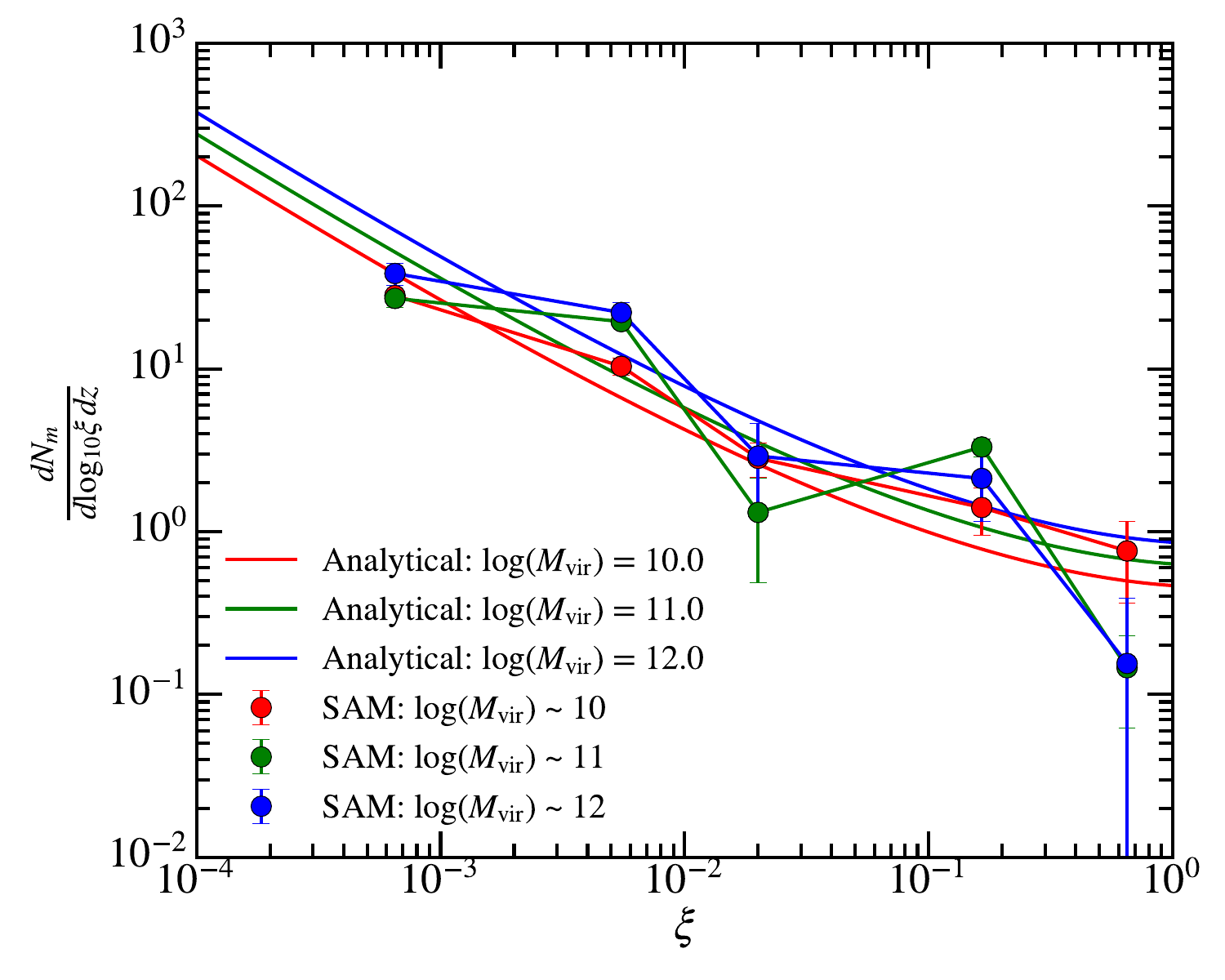}
    \caption{
        \textbf{Halo merger rates as a function of the mass ratios of the merging halos, $\xi$, at $z=6.43$.} We compare the merger rates obtained from our SAM (markers with error bars) and analytical merger rates calibrated using large-volume numerical simulation (solid lines) in three halo mass bins. We find an overall good agreement with analytical fittings despite statistical noises from the merger tree sampling.
    }
    \label{fig:merger_bootstrap}
\end{figure}

\section{Halo mass functions and halo merger rates}
\label{appsec:hmf}

The halo mass function describes the comoving number density of halos as a function of mass and redshift. As described above, we use the \texttt{HMF} package~\citep{Murray2013} to compute the halo mass functions analytically. In Fig.~\ref{fig:hmf_comparison}, we validate that the empirical weighting and counting scheme in Eq.~\ref{eq:hmf-tree} can reproduce the analytical halo mass functions at higher redshifts with reasonable accuracy.

The halo merger rate is a more important quantity to reproduce. To measure the merger rates in our SAM, we tracked halo evolution in discrete time steps of $\Delta z$, identifying progenitor haloes and their parent haloes at different redshifts. A merger event was defined as the case where two or more progenitor haloes at redshift $z+\Delta z$ merged to form a single descendant halo at redshift $z$. The mass ratio is $\xi = M_2 / M_1$, where $M_1$ and $M_2$ are the masses of the largest and the second-largest progenitor. We compute the merger event number density per unit redshift, $\frac{{\rm d}N_{\rm m}}{{\rm d}z}$, and normalize it by the halo mass function ($\Phi(M_1+M_2)$) to obtain the merger ``rate''.

To verify the robustness of our merger rates, we performed a consistency check against the fitting results from cosmological N-body simulations~\citep{Fakhouri2010}, where the halo merger rate per unit redshift per logarithm interval of progenitor mass ratio is parameterized as
\begin{equation}
\frac{{\rm d}N_{\rm m}}{{\rm d}\log \xi \, {\rm d}z} = \ln{10}\, A\,\left( \frac{M}{10^{12} \msun} \right)^{\alpha}\, \xi^{\beta + 1}\, e^{(\xi / \xi_{\star})^{\gamma}} \,(1 + z)^{\eta},
\label{hmreq}
\end{equation}
where $M$ is the descendant halo mass, $\xi$ is the progenitor mass ratio, and the best-fit parameters are~\citep{Fakhouri2010} $(\alpha, \beta, \gamma, \eta) = (0.133, -1.995, 0.263, 0.0993)$ and $(A, \xi_{\ast}) = (0.0104, 9.72 \times 10^{-3})$. The comparison is shown in Fig.~\ref{fig:merger_bootstrap}.

\section{SMBH mass functions and LISA predictions in model variations}

\begin{figure}[h!]
    \centering
    \includegraphics[width=\linewidth]{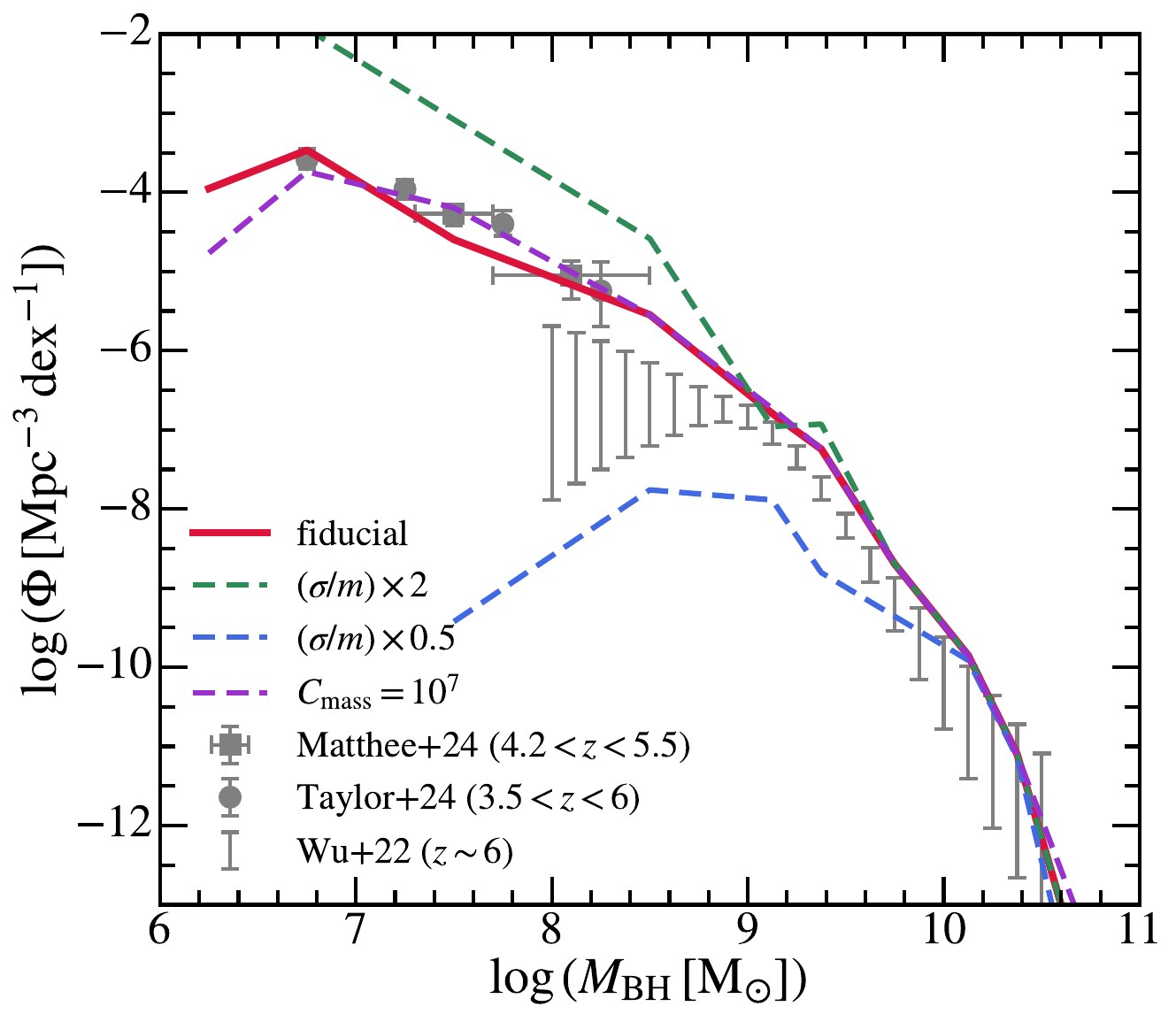}
    \caption{SMBH mass functions in models with different choices of $\sigma/m$. We experiment with increasing/decreasing $\sigma/m$ by a factor of two compared to the fiducial model, assuming the same $f_{\rm duty}$ as in the main text. The predicted mass function in the massive end is not affected. In the low-mass end, a lower $\sigma/m$ will prevent low-mass halos from collapsing to SMBH seeds, which results in a cut-off in the SMBH mass function. The implied $f_{\rm duty}$ will thus change substantially to match the observational constraints.}
    \label{appfig:bhmf}
\end{figure}

In this section, we explore how key model uncertainties influence our predictions for the SMBH mass functions and LISA detectability. We focus on two sources of uncertainty. The first concerns the dSIDM interaction cross-section, $\sigma/m$, which is degenerate with the free parameter $\epsilon$ in the collapse criterion. Varying either leads to similar effects. We investigate cases where $\sigma/m$ is increased/decreased by a factor of two. The resulting SMBH mass functions are shown in Fig.~\ref{appfig:bhmf} assuming the same AGN duty cycle as in the fiducial model. For the model with a twofold decrease in $\sigma/m$, dSIDM collapses into SMBHs only within more massive halos, leading to a stronger cutoff at the low-mass end. Although increasing the duty cycle $f_{\rm duty}$ could improve agreement with the quasar mass function from \cite{Wu2022}, this model still fails to account for the observed abundance of LRDs even with $f_{\rm duty}=100\%$. On the other hand, in the model with a twofold increase in $\sigma/m$, the massive end of the mass function remains unchanged, but more low-mass SMBHs ($M_{\rm BH} \lesssim 10^9,\msun$) are produced due to the increased dSIDM interaction rates. To match observations in this case, the duty cycle for low-mass AGN would need to be suppressed to sub-percent levels, close to what has been implied for massive quasars. Nevertheless, across all models considered, the predicted abundance of massive quasars remains unaffected. 

In addition, we also explore the potential impact of baryonic accretion using the merger-driven model developed in \cite{Volonteri2003,Volonteri2008,Xiao2021}. In this model, efficient gas inflow triggered by galaxy mergers feeds both the accretion of SMBHs and the star-formation in galaxy bulges. Following \cite{Volonteri2008}, we assume this feeding happens when the mass ratio between the two progenitor halos is larger than $0.1$ (defined as ``major merger''). The total amount of mass accreted during each major merger event is governed by the complicated gas dynamics and feedback processes in the galaxy bulge. Hypothetically, it manifests as the observed correlation between SMBH mass and bulge velocity dispersion of its host galaxy (i.e. the $M_{\rm BH}-\sigma_{\ast}$ relation \cite{Ferrarese2002,Kormendy2013})
\begin{equation}
    M_{\rm BH} = (4.4 \pm 0.9) \times 10^7 \msun \, (\sigma_{\ast}/150\kms)^{4.58\pm 0.52}.
\end{equation}
Motivated by this, we set the mass gain of an SMBH through accretion of baryonic matter during each merger event as
\begin{equation} 
    \Delta M_{\rm BH} = C_{\rm mass} (1-\epsilon_{\rm r}) \, (\sigma_{\ast}/150\kms)^{4.58},
    \label{eq:dmbh-sigma}
\end{equation}
where $\sigma_{\ast}$ is the bulge velocity dispersion of the merged galaxy, $\epsilon_{\rm r}$ is the radiative efficiency (assumed to be the canonical value $0.1$) and $C_{\rm mass}$ is a free normalization parameter, which has been set to $\sim 10^{4} \operatorname{-} 10^{7}\msun$ in previous studies of canonical seeds \cite{Volonteri2003,Volonteri2005,Volonteri2008,Natarajan2014}. In observations, at least in the local Universe, the bulge velocity dispersion is found to correlate with the asymptotic value of the halo circular velocity as \cite{Ferrarese2002}
\begin{equation}
    \log{V_{\rm c}} = (0.892 \pm 0.041) \log{\sigma_{\ast}} + (0.44\pm 0.09),
    \label{eq:Vc-sigma}
\end{equation}
where we approximate $V_{\rm c}$ using the maximum circular velocity of an NFW halo. The equations above result in a link between $\Delta M_{\rm BH}$ and host halo parameters ($M_{\rm vir}$, $c_{\rm vir}$) at a given redshift. This forms an empirical prescription to model the mass growth of SMBHs during galaxy mergers tracked by halo merger trees, with the assumption that the statistical correlations between SMBHs and their host galaxies (halos) are maintained throughout cosmic time. We refer the reader to \cite{Xiao2021} for details of this model. 

Here, to explore the maximum impact of baryon accretion, we make an aggressive choice of the normalization of the mass accretion per merger $C_{\rm mass}=10^{7}$, which will overshoot the local $M_{\rm BH}-\sigma_{\ast}$ relation. The SMBH mass function in this scenario is also shown in Fig.~\ref{appfig:bhmf}, which is almost indifferent from the fiducial model except for the mild shift of the cut-off in the low-mass end. Therefore, we conclude that baryonic accretion does not have a strong impact on these dissipative SIDM-seeded SMBHs, as most of the mass budget has a non-baryonic origin.

\begin{figure*}
    \centering
    \subfloat[$\sigma/m = 0.05\cpm$]{
        \includegraphics[width=0.32\textwidth]{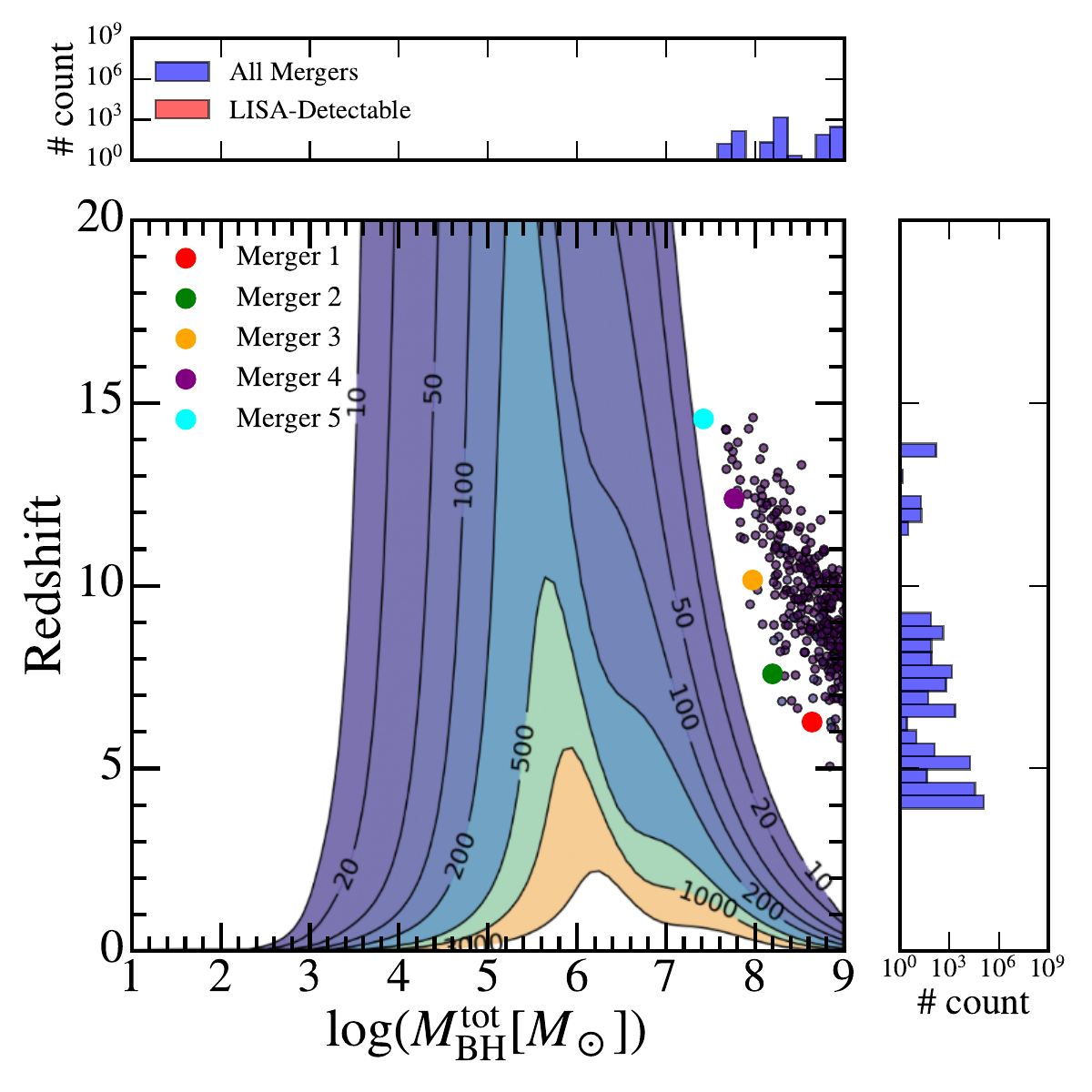}
    }
    \subfloat[$\sigma/m = 0.1\cpm$]{
        \includegraphics[width=0.32\textwidth]{plots/lisa_plots/lisa_single_merger_figure5.pdf}
    }
    \subfloat[$\sigma/m = 0.2\cpm$]{
        \includegraphics[width=0.32\textwidth]{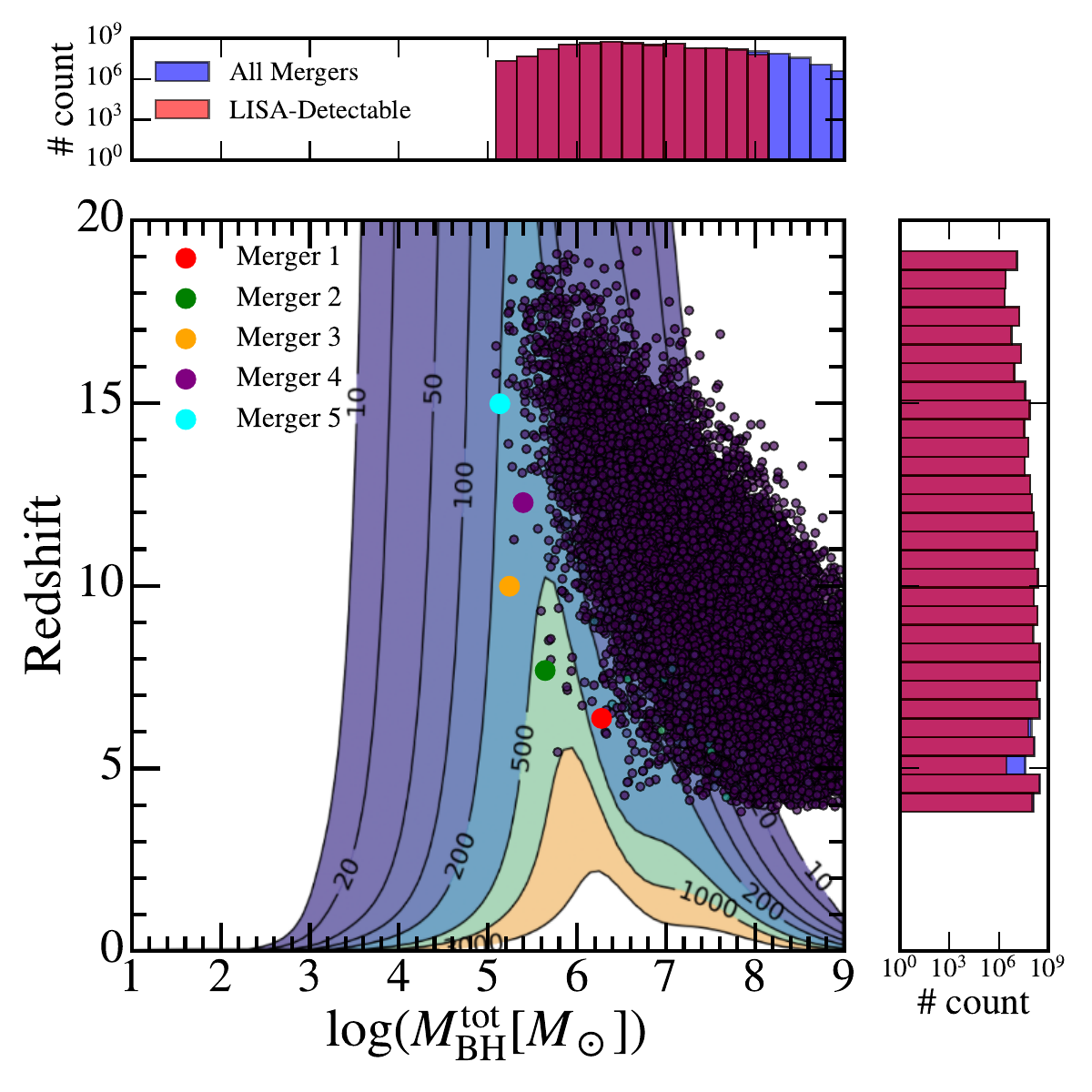}
    }
    \caption{
    LISA detectability of dSIDM-seeded SMBH mergers across three values of the self-interaction cross-section. The color-coded contours show the LISA SNR as a function of total SMBH mass $M_{\rm BH}^{\rm tot}$ and redshift, computed using the \texttt{PhenomD} waveform model. Selected mergers across different redshifts are marked on each plot. At low cross-section, mergers are dominated by massive SMBHs outside LISA's sensitivity band, while higher cross-sections shift the population to $\log(M_{\rm BH}^{\rm tot}/\msun)\sim 5$--$7$ where LISA is most sensitive.
    }
    \label{appfig:lisa_cmass0}
\end{figure*}

\begin{figure*}
    \centering
    \subfloat[$\sigma/m = 0.05\cpm$]{
        \includegraphics[width=0.32\textwidth]{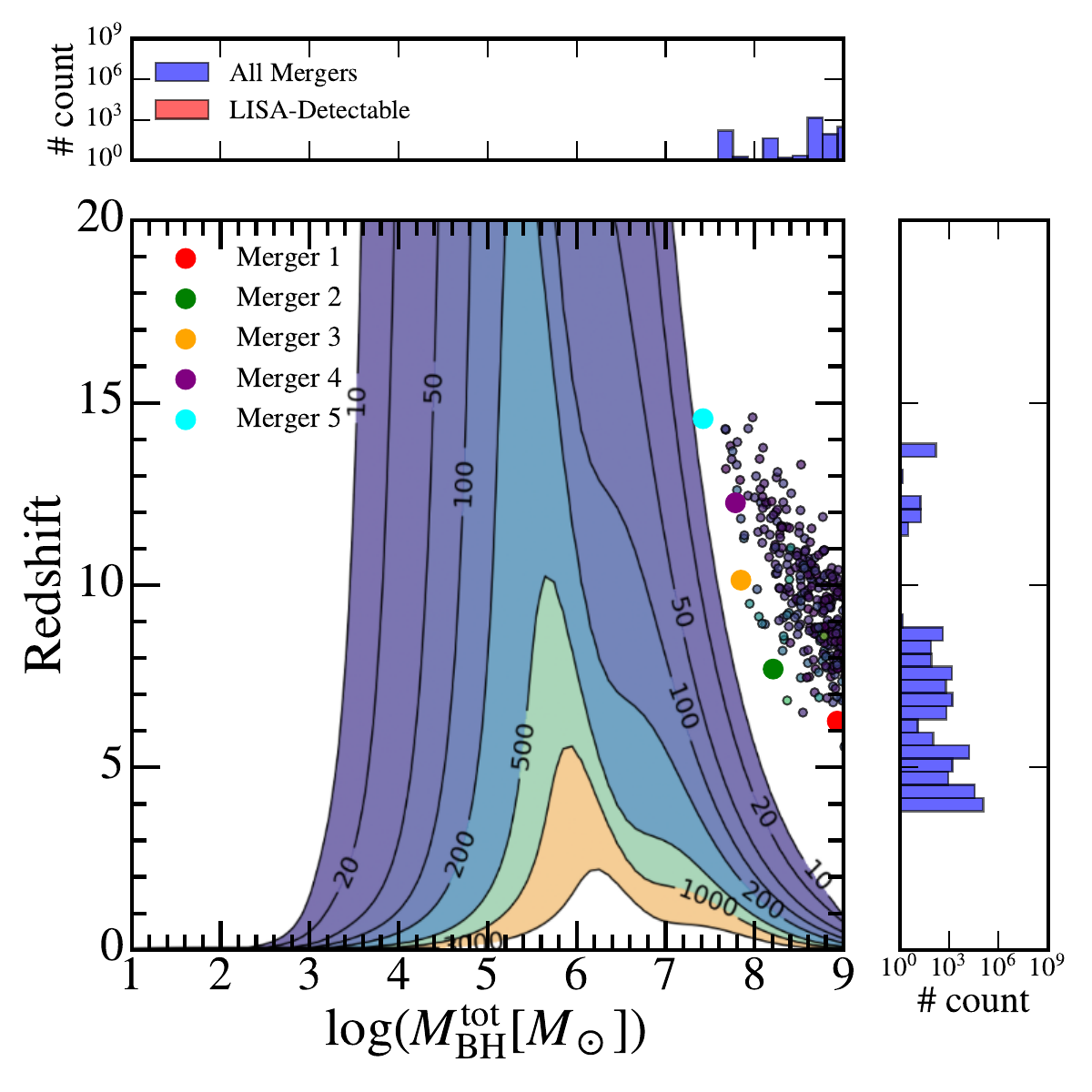}
    }
    \subfloat[$\sigma/m = 0.1\cpm$]{
        \includegraphics[width=0.32\textwidth]{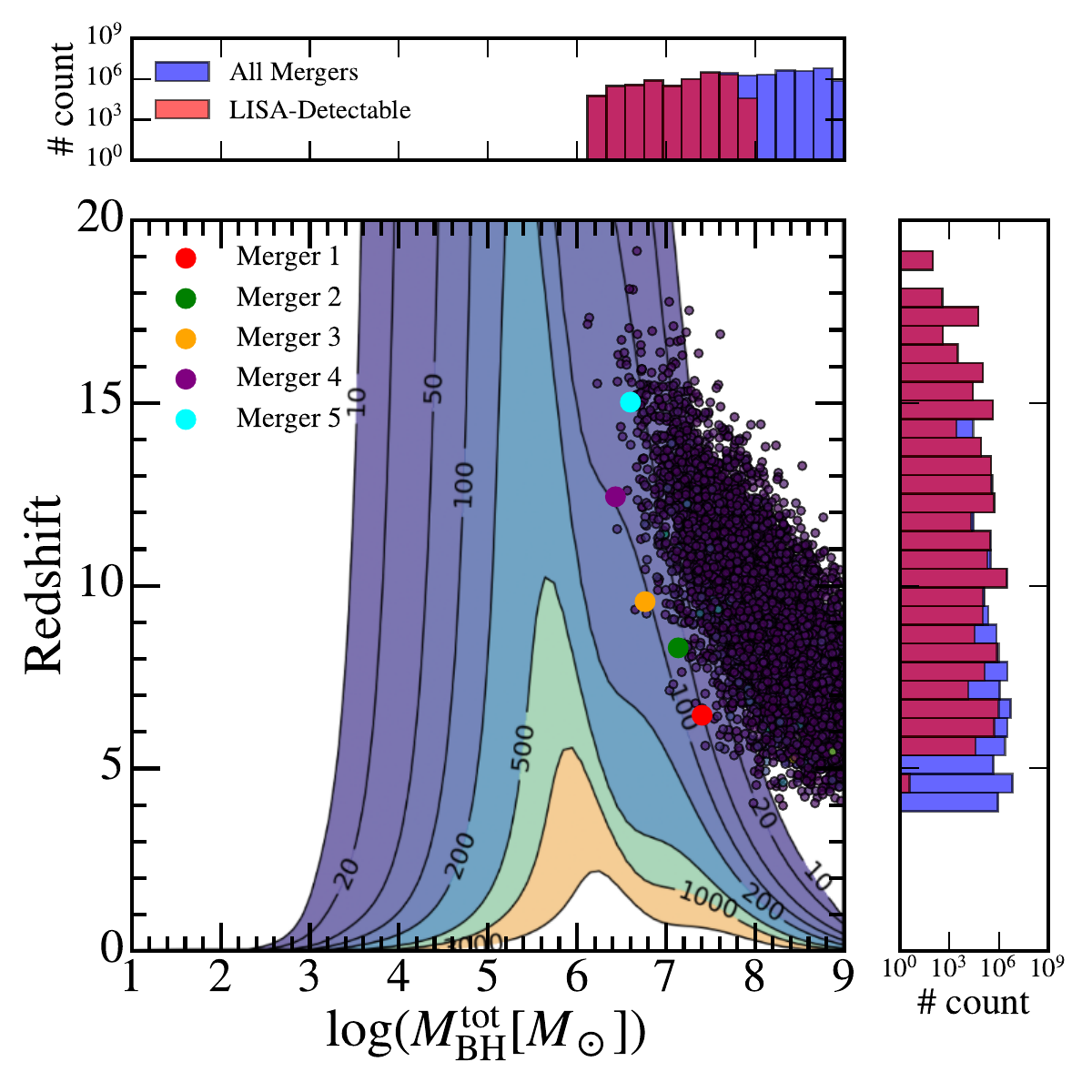}
    }
    \subfloat[$\sigma/m = 0.2\cpm$]{
        \includegraphics[width=0.32\textwidth]{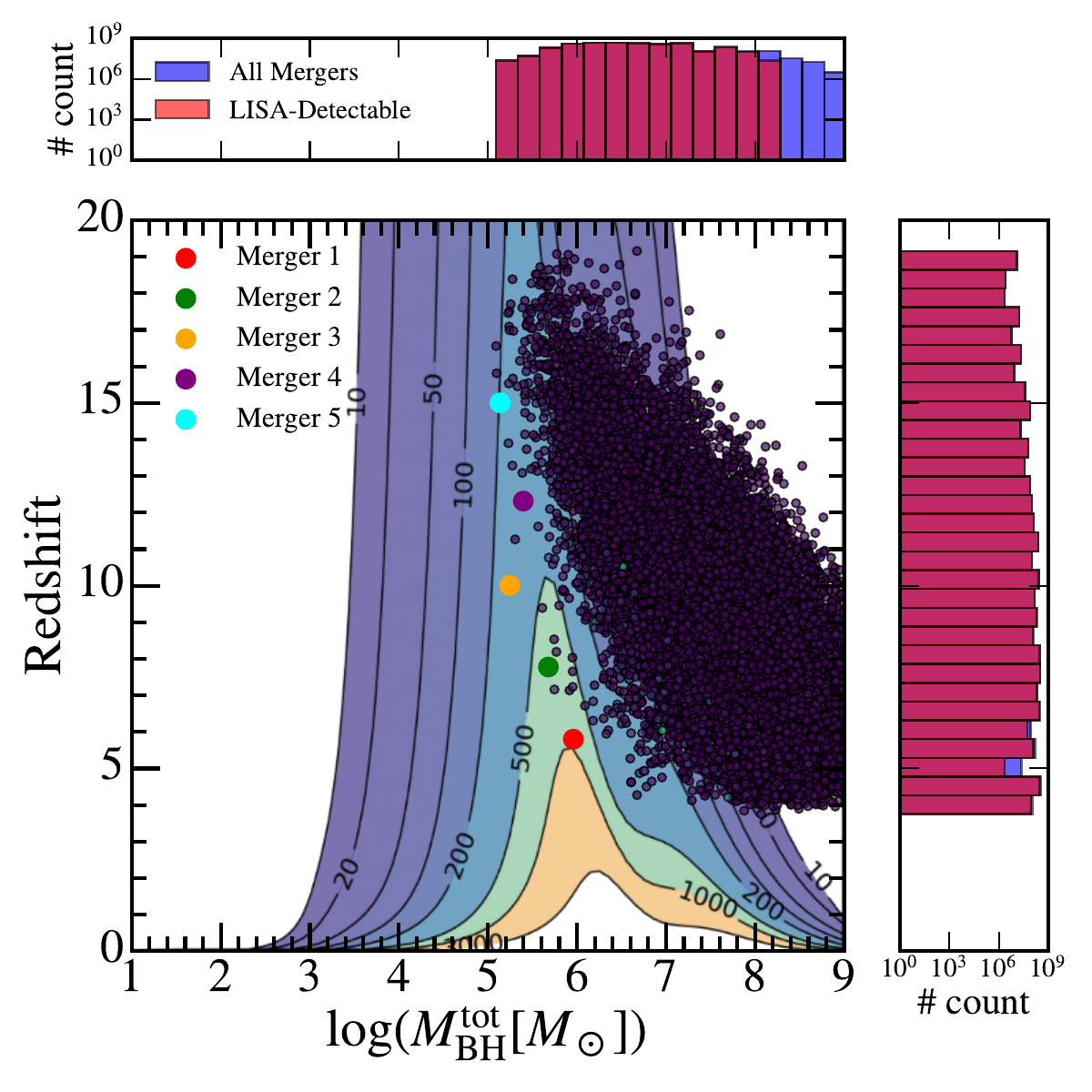}
    }
    \caption{
    Same as Fig.~\ref{appfig:lisa_cmass0} but with baryonic accretion included ($C_{\rm mass} = 10^7$).
    }
    \label{appfig:lisa_cmass1e7}
\end{figure*}

The LISA detectability of SMBH mergers across these cross-section model variations is shown in Fig.~\ref{appfig:lisa_cmass0}. Since $t_{\rm col} \propto (\sigma/m)^{-1} M_{\rm vir}^{-1/3}$ (Eq.~1 in the main text), a higher $\sigma/m$ extends the seeding criterion to lower-mass halos. For $\sigma/m = 0.05\cpm$, only massive halos host dSIDM-seeded BHs, and the typical binary total BH mass is $\log(M^{\rm tot}_{\rm BH}/\msun) \gtrsim 8$, which lies outside LISA's detectable band. For the fiducial and higher cross-sections ($\sigma/m = 0.1$ and $0.2\cpm$), seeding extends to $M_{\rm vir} \sim 10^{9}$--$10^{10}\msun$, shifting merger BH binaries to $\log(M^{\rm tot}_{\rm BH}/\msun) \sim 5$--$7$ where LISA is most sensitive. In Fig.~\ref{appfig:lisa_cmass1e7}, we show the prediction when baryonic accretion is included, assuming $C_{\rm mass}=10^{7}$. Similar to our findings in the SMBH mass function above, baryonic accretion has little impact on the predictions of SMBH merger events.

\end{document}